\documentclass[authoryear,preprint,review,12pt]{elsarticle}

\usepackage{subfigure}
\usepackage{setspace}
\usepackage{geometry}
\usepackage{framed,multirow}
\usepackage{float}
\usepackage{latexsym}
\usepackage{bm}
\usepackage{amsmath}
\usepackage{booktabs}
\usepackage{xcolor}
\usepackage{array}
\usepackage{algorithmic}
\usepackage{dblfloatfix}
\usepackage{verbatim}
\usepackage{mathrsfs}
\usepackage{amsthm,amsmath,amssymb}
\usepackage{lineno,hyperref}

\usepackage{amssymb}

\begin{document}

\makeatletter
\def\ps@pprintTitle{%
  \let\@oddhead\@empty
  \let\@evenhead\@empty
  \def\@oddfoot{\reset@font\hfil\thepage\hfil}
  \let\@evenfoot\@oddfoot
}
\makeatother

\begin{frontmatter}



\title{Information fusion approach for biomass estimation in a plateau mountainous forest using a synergistic system comprising UAS-based digital camera and LiDAR}


\author{
 Rong Huang\textsuperscript{a}, Wei Yao\textsuperscript{a,b*}, Zhong Xu\textsuperscript{c}, Lin Cao\textsuperscript{c}, Xin Shen\textsuperscript{c}}

\address{
	\textsuperscript{a}Department of Land Surveying and Geo-Informatics, The Hong Kong Polytechnic University, Hung Hom, Hong Kong\\
	
	\textsuperscript{b}The Hong Kong Polytechnic University Shenzhen Research Institute, Shenzhen, China\\
	
	\textsuperscript{c}Co-Innovation Center for Sustainable Forestry, Nanjing Forestry University, 159 LongpanRoad, Nanjing, 210037, China\\
	
}

\begin{abstract}
Forest land plays a vital role in global climate, ecosystems, farming and human living environments. Therefore, forest biomass estimation methods are necessary to monitor changes in the forest structure and function, which are key data in natural resources research. Although accurate forest biomass measurements are important in forest inventory and assessments,
high-density measurements that involve airborne light detection and ranging (LiDAR) at a low flight height in large mountainous areas are highly expensive. The objective of this study was to quantify the aboveground biomass (AGB) of a plateau mountainous forest reserve using a system that synergistically combines an unmanned aircraft system (UAS)-based digital aerial camera and LiDAR to leverage their complementary advantages. In this study, we utilized digital aerial photogrammetry (DAP), which has the unique advantages of speed, high spatial resolution, and low cost, to compensate for the deficiency of forestry inventory using UAS-based LiDAR that requires terrain-following flight for high-resolution data acquisition. Combined with the sparse LiDAR points acquired by using a high-altitude and high-speed UAS for terrain extraction, dense normalized DAP point clouds can be obtained to produce an accurate and high-resolution canopy height model (CHM). Based on the CHM and spectral attributes obtained from multispectral images, we estimated and mapped the AGB of the region of interest with considerable cost efficiency. Our study supports the development of predictive models for large-scale wall-to-wall AGB mapping by leveraging the complementarity between DAP and LiDAR measurements.
This work also reveals the potential of utilizing a UAS-based digital camera and LiDAR synergistically in a plateau mountainous forest area.


\end{abstract}



\end{frontmatter}


\section{Introduction}
As the predominant terrestrial ecosystem on earth, forests influence regional and global climate change, shape human living environments and even help cultivate the farming land. Therefore, they are very important in environment and natural resources management \citep{givoni1998climate,pan2011large}.
Concurrently, forest trees, being important carbon sinks and pools, significantly impact global carbon dynamics \citep{pan2011large}.  
Biomass is a important factor for quantifying carbon storage in forest vegetation because it indicates the biological and geometrical characteristics of trees. 
Crown volume, constituting more than 30\% of the total biomass in coarse and fine branches, leaves, and fruits, is another important characteristic for tree biomass estimation \citep{goodman2014importance,ploton2016closing}.
Therefore, accurate forest biomass estimation is crucial for comprehending the role of forests in the global carbon cycle. 
However, biomass is one of the most difficult tree parameters to estimate, especially because of the difficulty in capturing irregularly shaped crowns through standard forest fieldwork such as manual surveys \citep{le2019new}. To achieve highly accurate forest biomass estimation of large areas, heavy ground-based measurements of inventory plots are required. Many natural forest areas are remote/sparsely populated, limiting the feasibility of field surveys. Remote sensing data are therefore a practical choice to provide data support for biomass estimation, especially in large forest areas. 

Many researchers have utilized data obtained from remote sensors, such as the Landsat and MODIS systems, to estimate forest cover and carbon stocks from local to global scales, as well as changes in forest extent and biomass density \citep{aide2013deforestation, brosofske2014review, matasci2018large}. The spectral information provided by these optical images can be used to identify tree species that are important for precise biomass estimation. 
However, images from satellites or airplanes usually have coarse spatial resolution.
Moreover, remote sensing images require calibration and correction because of the effects of aerosols and water vapor. 
An unmanned aerial system (UAS) with multiple sensors could be useful for acquiring ultrahigh spatial resolution data, thereby boosting forest inventory.


As an alternative to optical remote sensors, light detection and ranging (LiDAR) can provide useful structural information on forests for biomass estimation. 
Many researchers have reported methods for obtaining various forest variables based on LiDAR measurement \citep{wang2016international,yu2017single}. 
Specifically, in a wide variety of forestry applications, LiDAR point clouds are often used as a data source for retrieving forest variables, including stem density \citep{richardson2011strengths}, tree height, tree species, crown size, wood volume \citep{yao2012tree}, and aboveground biomass (AGB) \citep{strimbu2017post}. 
Specifically, airborne laser scanning data can be employed to obtain three-dimensional structures of trees, facilitating efficient large-scale 3D forest inventory \citep{yao2012tree} and providing practical solutions to the challenges posed by large-scale biomass estimations in forest areas. 

AGB estimation methods can be classified into two major categories. The first is individual-tree-based methods. 
In these methods, individual trees are detected, isolated, and segmented from the measured point clouds. Subsequently, the tree biomass was modeled by calculating the volume of the tree branches and stems through 3D geometric reconstruction of the tree structure. Many researchers have focused on individual tree crown segmentation \citep{li2012new} and tree species classification \citep{yao2012tree, xu2020tree}.  
The easiest tree modeling method is plot-scale inventory using predefined regression models such as the diameter at breast height (DBH) \citep{yao2012tree} and height \citep{dassot2012terrestrial}. 
In addition, several researchers have focused on developing automated modeling methods or algorithms for generating a detailed representation of single trees. One method entails fitting regular geometric models, such as circles \citep{bienert2007tree} and cylinders \citep{thies2004three}, using voxel-based \citep{jupp2009estimating} or mesh-based representations \citep{antonarakis2009leafless}. 
The model-fitting-based method using regular geometric models, particularly the parametric model, can be used directly for tree volume estimation; hence, it is an easy solution.
However, real tree structures are randomly formed and the difference between regular geometric models and actual shapes of tree structures affects the accuracy of volume estimation \citep{le2019new}.
The second is plot-based methods, in which, instead of individual tree detection, the LiDAR measurements of canopy surface height can be matched to field-measured tree inventory plots with known biomass. 
Thus, a predictive model can be developed based on the reference biomass estimates and applied to large-scale areas \citep{asner2014mapping}. 
Different structural metrics based on LiDAR measurements were extracted for the large-area mapping of the Canadian boreal forest biomass in \cite{matasci2018large}.

For areas with high relief and undulated terrain, aboveground forest biomass estimation is difficult because of the difficulty in acquiring precise data of equal quality \citep{soenen2010estimating}. Conventional methods focus on retrieving spectral information for estimating parameters but ignore the subcanopy structure; therefore, they cannot estimate biomass with high precision. However, as mentioned in a previous review, the accuracy of the biomass estimation corresponds to that of the tree height measurement. Although the UAS-based LiDAR can solve this problem, it is resource-intensive and time-consuming.

\begin{figure}[htb]
    \centering
    \includegraphics[width=\textwidth]{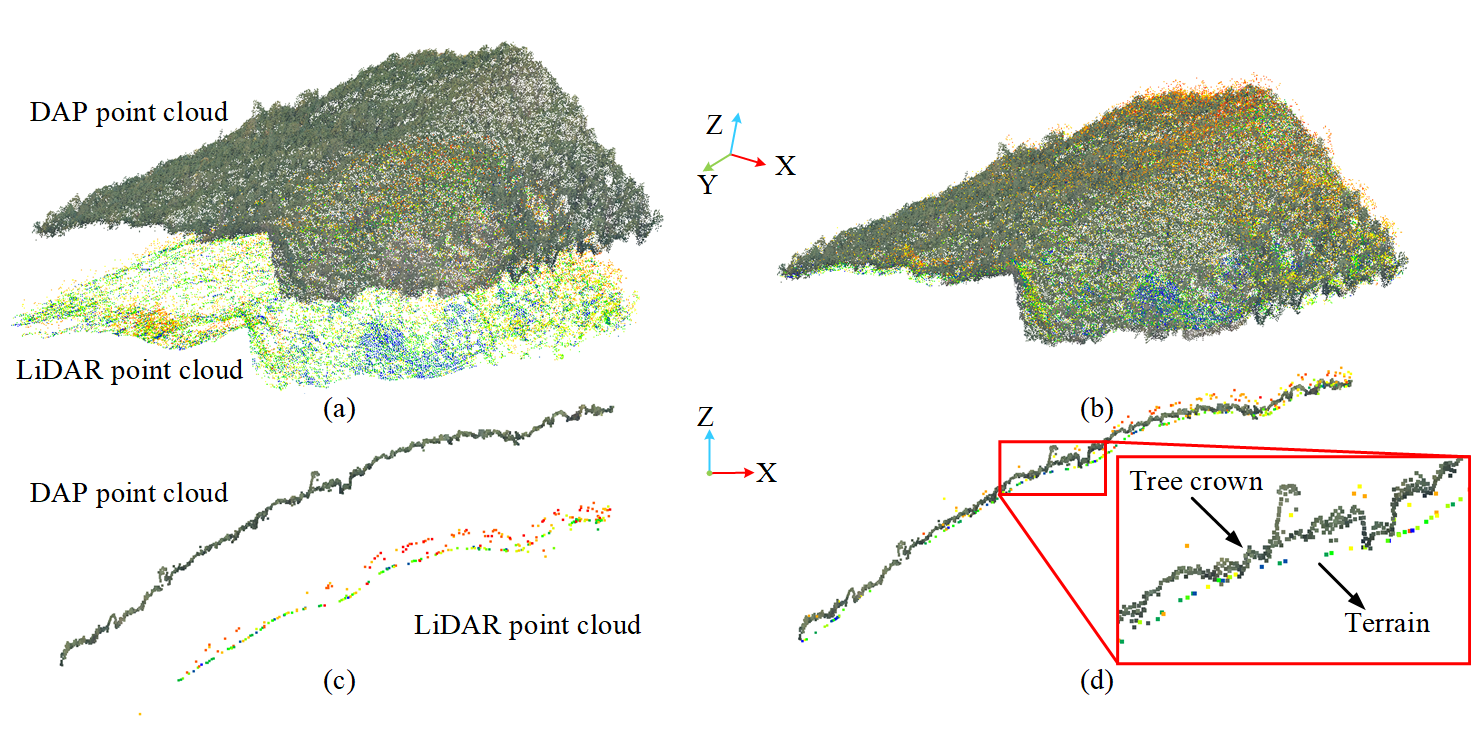}
    \caption{Acquired DAP and LiDAR point clouds of a sample area. a) DAP and LiDAR point clouds before registration, b) DAP and LiDAR point cloud after registration, c) cross-section of a, and d) cross-section of b. DAP point cloud is colored by RGB, and LiDAR point clouds are colored by intensities.}
    \label{fig:example_illustration}
\end{figure}

In this study, we aimed to measure the biomass of a plateau mountainous forest reserve. 
In such areas, terrain-following, low-speed flights substantially increase the cost and challenges of acquiring high-quality LiDAR data. 
Furthermore, acquiring LiDAR data with equal quality is challenging and certain flight parameters must be adapted to the terrain of the measured area. 
This requires precise control of the flight route, which significantly increases the data acquisition cost. 
This runs counter to our primary intent of using remotely sensed data with high efficiency for large-scale biomass estimation. 
Digital aerial photogrammetry (DAP) has the unique advantages of speed, high spatial resolution, and low cost, which could compensate for the deficiency of conventional methods. 
However, as shown in Fig.~\ref{fig:example_illustration}, DAP technology can only measure the canopy of trees; it tends to underestimate the tree structure below the canopy. 
Although LiDAR point clouds acquired using high-speed UAS cannot yield accurate measurements of tree structures, they are effective for terrain prediction. 
Thus, a combination of DAP point clouds and sparse LiDAR point clouds would efficiently mitigate the disadvantages of both data sources and accurately extract the point cloud structural attributes. This system eliminates the need for terrain-following flight when acquiring high-quality LiDAR data and enables efficient low-cost acquisition of high-quality data for biomass estimation. Using the structural and spectral attributes obtained from the point clouds and multispectral imagery, respectively, we developed a predictive model for biomass estimation and applied it to the region of interest. The major objectives of this study were as follows: (i) to generate normalized DAP point clouds by combining dense DAP point clouds with spare LiDAR point clouds for accurately estimating canopy height model (CHM); (ii) to develop a predictive model for estimating forest biomass based on the spectral and point cloud structural attributes; and (iii) to accurately estimate the biomass of a large plateau mountainous forest area.
The major contributions of this study can be summarized as follows:
\begin{itemize}
    \item A synergistic system that leverages the complementary integration of UAS-based digital camera and LiDAR sensors was established for AGB estimation in a plateau mountainous forest area to optimize the trade-off between  efficacy and cost/risk for data acquisition.
    \item A hierarchical pipeline that can exploit the advantages of high-resolution aerial imagery and sparse LiDAR point clouds using a global information fusion approach and compensate for their defects was proposed for AGB estimation,.
    \item We achieved biomass mapping in a large-scale forest area located in a plateau mountainous but subtropical landscape with high vegetation coverage and analyzed the effect of using multisource data on AGB mapping.
\end{itemize}





\section{Materials and methods}

\subsection{Study site}
\begin{figure}[htb]
    \centering
    \includegraphics[width=\textwidth]{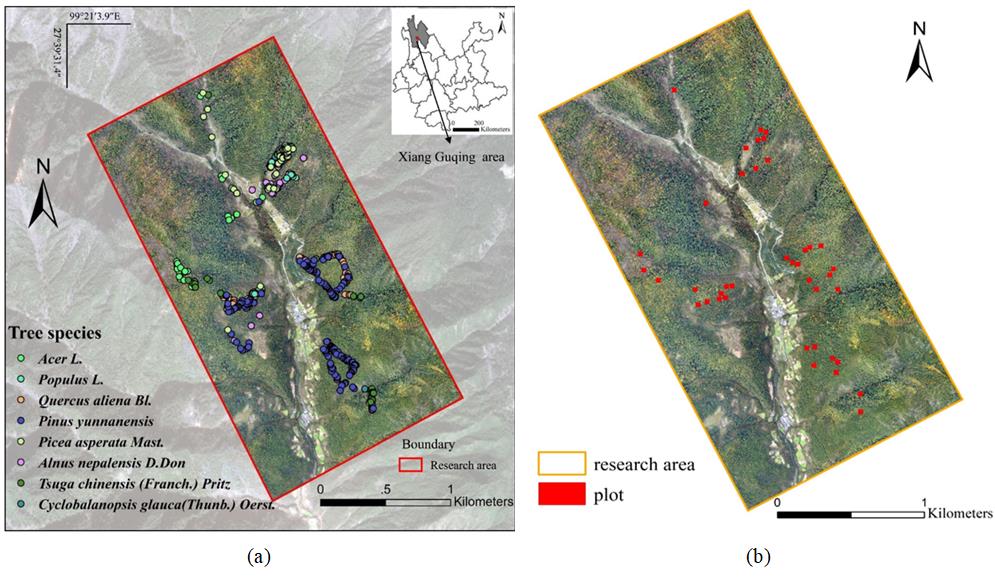}
    \caption{Overview of the XGQ study site. a) Illustration of the distribution of different tree species, as determined by fieldwork \citep{xu2020tree} and b) distribution of selected sample plots for developing models for AGB estimation.}
    \label{fig:study_area}
\end{figure}

Xiangguqing (XGQ) is in the northeast of Yunnan Province and the southern end of the Baima Snow Mountain National Nature Reserve ($99^\circ21'3.9''$E, $21^\circ39'14''$N). Owing to the subtropical and highland monsoon climate of the northern hemisphere, it is home to a large natural forest area dominated by two main species, conifer and broadleaf species. Climate in this study indicates temperature, precipitation, and species distribution patterns. Specifically, the mean annual rainfall and temperature are approximately 1370.7 $mm$ and 9.8 $^\circ$C, respectively. The study site is approximately 1000-ha area, with an elevation ranging from 2300-3500 $m$. It is dominated by dense forests on valleys with steep slopes and some villages and farmlands (see Fig.~\ref{fig:study_area}). Thirty four sample plots with a side length of 30 m were selected for modeling the AGB estimation in the study area, which was utilized for developing a predictive model for AGB estimation in the forest area. 

\subsection{Field and remote sensing data}
From October to December 2018, field and remote sensing data were collected by using UASs and conducting field surveys \citep{xu2020tree}. As mentioned earlier, the dominant tree species in XGQ can be classified into two categories: (i) conifer species: Pinus yunnanensis (P.Y.), Picea asperata (P.A.), and Tsuga chinensis (T.C.); and (ii) broadleaf species: Cyclobalanopsis oxyodon (C.O.), Acer forrestii (A.F.), Populus davidiana (P.D.), Quercus aliena (Q.A.), and Alnus nepalensis (A.N.).

Fieldwork was conducted in the study area from October 29 to November 4, 2018. In the field survey, a total of 502 trees were measured. The data collected included tree positions, species, DBH, treetop heights, and crown widths in both cardinal directions. To provide representative samples, the sizes of trees, including the dominant tree species and other trees in the reserve, were measured. At least 15 samples were analyzed for each tree species. The measured trees were distributed spatially evenly throughout the reserve. The positions of individual trees were measured by using a Trimble® real-time kinematic (RTK) R4 GNSS receiver with centimeter-level accuracy. The DBH of the trees was measured using diameter tape. The treetop heights were measured by using a Vertex IV hypsometer. The average of the two values measured in the two vertical directions was calculated to obtain the crown widths. Table~\ref{tab:tree_specices} presents the statistics for the eight dominant tree species based on three major tree parameters.
\begin{table*}[ht!]	
    \centering
    \caption{Summary of the statistics of the measured trees, including the mean value and standard deviation of DBH, height, and crown width. \label{tab:tree_specices}}
    \resizebox{0.9\textwidth}{!}
    {\begin{tabular}{c c c c c c c c}
    \toprule
       \multirow{2}{*}{Tree species}  & \multirow{2}{*}{N} & \multicolumn{2}{c}{DBH ($cm$)} & \multicolumn{2}{c}{Height ($m$)} & \multicolumn{2}{c}{Crown width ($m$)}\\
       & & Mean & SD & Mean & SD & Mean & SD\\
       \midrule
       Pinus yunnanensis (P.Y.) & 162 & 36.26 & 15.83 & 20.13 & 12.03 & 6.20 & 1.94\\
       Picea asperata (P.A.) & 39 & 53.95 & 18.90 & 28.67 & 10.37 & 7.47 & 2.43\\
       Tsuga chinensis (T.C.) & 46 & 57.97 & 27.73 & 24.14 & 7.90 & 8.51 & 2.98\\
       Cyclobalanopsis oxyodon (C.O.) & 59 & 53.37 & 35.84 & 20.55 & 7.09 & 7.25 & 3.02\\
       Acer forrestii (A.F.) & 32 & 55.83 & 50.18 & 17.70 & 5.30 & 8.15 & 2.71\\
       Populus davidiana (P.D.) & 30 & 27.04 & 8.24 & 16.72 & 4.08 & 5.21 & 2.31\\
       Quercus aliena (Q.A.)& 83 & 34.83 & 13.53 & 15.71 & 3.83 & 6.90 & 2.52\\
       Alnus nepalensis (A.N.) & 51 & 32.74 & 11.12 & 16.65 & 5.58 & 6.56 & 2.49\\
        \bottomrule
    \end{tabular}}
\end{table*}

\begin{figure}[ht!]
    \centering
    \includegraphics[width=0.6\textwidth]{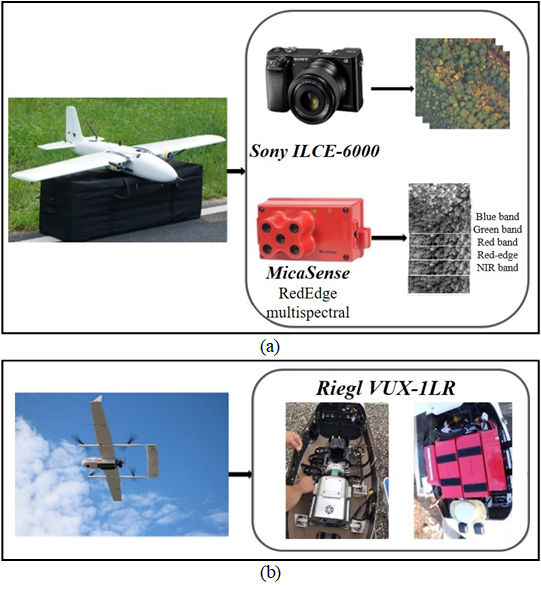}
    \caption{RGB UAS images and LiDAR system used in this study \citep{xu2020tree}. a) Fixed-wing UAS platform with multispectral and RGB sensors and b) CW-30 UAS with a LiDAR system.}
    \label{fig:data_acquisition}
\end{figure}

From October 29 to 30, 2018, RGB and multispectral images were acquired using a fixed-wing UAS mounted with Sony ILCE-6000 and MicaSense RedEdge multispectral sensors (see Fig.~\ref{fig:data_acquisition} (a). During the acquisition period, the weather was clear and partly cloudy. The parameters for the imagery acquisition are listed in Table~\ref{tab:data_dap}.
\begin{table*}[ht!]
    \centering
    \caption{Imagery acquisition parameters.\label{tab:data_dap}}
    \resizebox{\textwidth}{!}
    {\begin{tabular}{c c c}
    \toprule
       Parameters & Sony ILCE-6000 & MicaSense RedEdge\\
       \midrule
       Flight height ($m$) & 800 & 800\\
       Flight speed ($m/s$) & 20 & 30\\
       Forward overlap ($\%$) & 80 & 80 \\
       Lateral overlap ($\%$) & 70 & 80\\
       Spectral bands & Blue, green, red & Blue, green, red, red-edge, and near-infrared \\
       Optimal resolution  & 6000 $\times$ 4000 & 1280 $\times$ 960\\
       Ground sample distance ($m$)  & 0.1 & 0.5\\
       Image format & 24-bit TIFF & 16-bit TIFF \\
       \bottomrule
    \end{tabular}} 
\end{table*}

In addition to the optical data, LiDAR data were obtained using a CW-30 on a JOUAV series fixed-wing UAS with a Riegl VUX®-1 LR sensor ( Fig.~\ref{fig:data_acquisition} (b)). The fuselage length, wingspan, and maximum take-off weight of the UAS were 2.1 $m$, 4.0 $m$, and 34.5 $kg$, respectively. It was equipped with an 80-$cc$ carburetor engine. Positioning accuracy was achieved using real-time kinematic technology. The take-off and landing modes were vertical take-off and landing. The data were acquired on December 16, 2018. The parameters of the LiDAR data acquisition are listed in Table~\ref{tab:data_als}.
\begin{table*}[ht!]
    \centering
    \caption{LiDAR acquisition parameters.\label{tab:data_als}}
    \resizebox{0.55\textwidth}{!}
    {\begin{tabular}{c c}
    \toprule
       Parameters & Riegl VUX-1 LR\\
       \midrule
       Flight height ($m$) & 900\\
       Flight speed ($m/s$) & 27\\
       Strip spacing ($m$) & 400 \\
       Laser wave length ($nm$) & 1550\\
       Divergence ($mrad$) & 0.5 \\
       Pulse emission frequency ($KHz$) & 50-820\\
       Scanning frequency ($KHz$) & 100\\
       Point density ($pt/m^2$) & 0.627\\
        \bottomrule
    \end{tabular}}
\end{table*}

\subsection{Data preprocessing}
The RGB and multispectral images were processed to generate ortho-images with high resolution, rich spectral information, and DAP point clouds with high point densities. First, the optical images were registered using GPS data and IMU measurements, following which the aligned images were stitched to generate an ortho-mosaic \citep{goodbody2017unmanned}. 
Second, a spline algorithm \citep{vercauteren2009diffeomorphic} was used to convert multispectral images to RGB ortho-imagery using selected 200 ground control points. 
Consequently, the mean $RMSE$ for both the $X-$ and $Y-$ orientations was less than 0.01 $m$. 
Finally, the nearest neighbor pixel method was used for resampling to obtain spatially matched multispectral and RGB images with a spatial resolution of 0.1 $m$. 
Thus, ortho-imageries with rich spectral information were obtained.

In contrast, DAP point clouds with high point densities were obtained using the structure from motion (SfM) algorithm, following which the RGB images were aligned using positioning data. Then, the tie points in overlapping areas, as the skeleton of the point cloud, were generated \citep{goodbody2017unmanned}. 
The RGB-based DAP dense point clouds were then generated using the SfM technology. 
The final point density of the DAP point cloud was 42.45 $pt/m^2$.

\subsection{Overall workflow}

\begin{figure}[ht!]
    \centering
    \includegraphics[width=0.85\textwidth]{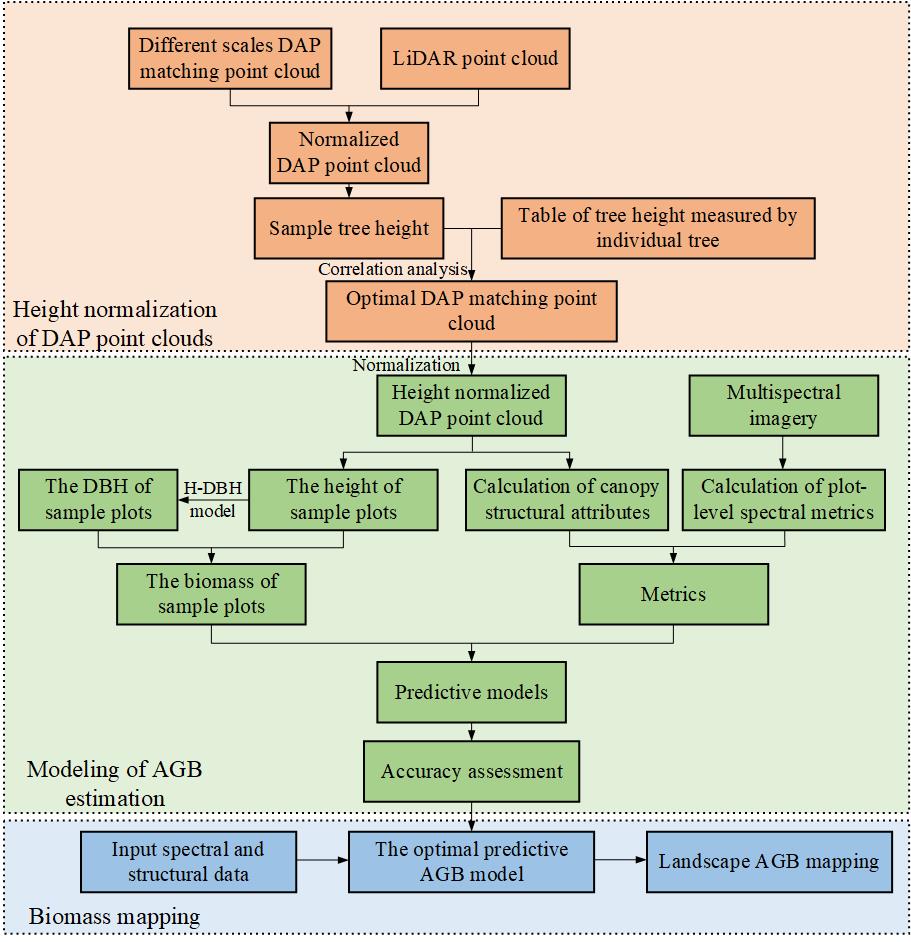}
    \caption{Overview of the processing framework.}
    \label{fig:method_workflow}
\end{figure}

The workflow consisted of three steps (see Fig.~\ref{fig:method_workflow}). 
First, the DAP point clouds were registered to the LiDAR point clouds to normalize the DAP point clouds with respect to the terrain measured by the LiDAR. This made it possible to extract the structural attributes of the trees by generating the CHM. 
Second, the tree crowns of the selected sample plots were segmented. 
Then, according to the results of the identification of individual tree crowns and tree species, the tree-level AGB of the sample plots was calculated based on the height-to-AGB models of the dominant tree species. 
Third, according to the estimated AGBs of the sample plots, a model was developed for plot-level AGB estimates using structural attributes from the CHM and plot-level spectral attributes extracted from multispectral data. 
Based on the predictive AGB model, we obtained the AGB-plot estimates of the target landscape. 
The detailed steps are explained in the subsequent sections.

\subsection{Height normalization of DAP point clouds with respect to LiDAR point clouds}
First, we normalized the DAP point clouds with respect to the LiDAR point clouds to generate the CHM.
As mentioned earlier, the estimated structural information of trees is important for the calculation of the AGB. 
Tree height is an important indicator for calculating most of the structural metrics. 
However, owing to canopy coverage, it is difficult to estimate AGBs from DAP point clouds. 
Moreover, LiDAR point clouds obtained on high flight speeds were too sparse to be utilized for accurately extracting structural information. 
In this study, we matched the DAP point cloud with LiDAR point cloud. 
The latter served as reference data for modeling the terrain under canopy coverage because the laser penetrated the canopy. 
As the optical images and LiDAR point clouds were acquired using different platforms and positioning data from GPS and IMUs were not sufficient for accurately aligning point clouds from these two data sources, it was necessary to match the point clouds. However, it was challenging, owing to variations in measured details (see Fig.~\ref{fig:example_illustration}). In previous research, most of the proposed marker-free registration methods in the field of forest monitoring were aimed at registering point clouds of similar quality, such as multi-scan terrestrial laser scanning point clouds \citep{guan2020marker,kelbe2016marker}, or point clouds from different sources registered using individual object attributes such as tree position \citep{polewski2019marker} or artificial linear structure \citep{POLEWSKI201979}. However, the data quality of the DAP point clouds varied significantly from that of the LiDAR point clouds. Further, individually extracting reliable tree-related parameters for registration was challenging. 
Thus, in this study, instead of matching local details, we introduced a global marker-free registration method \citep{huang2021robust} that utilized global information and was robust to noise and applicable to low-overlapping cases.
Subsequently, we first presented the detailed steps for normalization and then presented the key registration method.

The normalization can be divided into three sub-steps. 
First, the LiDAR strips were matched to generate a full-coverage LiDAR point cloud. 
The optimal parameters of the LiDAR point cloud registration were determined by evaluating the matching results using the sum of the point-to-point distance residuals and registration result with the lowest matching errors was selected as the optimal matching result. 
Second, the DAP point clouds were matched with the united LiDAR point cloud to generate a normalized DAP point cloud. 
The accuracy of multisource registration significantly influenced the CHM extraction. 
Matching results under different parameterizations were obtained. To estimate the optimal parameters used for matching DAP and LiDAR point clouds, registration with several different scales was conducted.
The registration results were evaluated using sample tree heights that were obtained using the normalized DAP point cloud and compared with the tree heights measured in fieldwork. A correlation analysis was performed to select the optimal matching result and use it to generate the final normalized point cloud.
Third, once the optimal matching result between the DAP and LiDAR point clouds had been obtained, the DAP point clouds could be normalized based on the ground points extracted from the LiDAR point clouds using the progressive triangulated irregular network densification filtering algorithm \citep{zhao2016improved}. The ground points were used to generate a digital terrain model (DTM) with a resolution of 0.5 $m$. Subsequently, a digital surface model was created using the points with the highest heights from the DAP point clouds and normalized with reference to the generated DTM. Finally, the normalized DAP point clouds were processed to generate a CHM with a spatial resolution of 0.1 $m$ corresponding to the spatial resolution of the fused optical images using the natural neighborhood interpolation algorithm. The registration method and correlation analysis are explained in the subsequent sections. 

As previously explained, to match point clouds from two different platforms, we adapted a point cloud registration method called GRPC \citep{huang2021robust} for the special case of aligning the DAP and LiDAR point clouds. As illustrated in Fig.~\ref{fig:example_illustration}, the measured details in both point clouds are considerably different, thus posing significant challenges for extracting local correspondences for matching. However, as the variation in tree heights is less pronounced than the undulation of mountain terrain, the general trend of the canopy surface coincides with the trend of terrain under the canopy. To match the DAP and LiDAR point clouds, we utilized the information on global trends provided by the two data sources. Thus, the GRPC was introduced as a matching strategy that utilized the spatial distribution information and matched the global correspondences with operations in the frequency domain.

The GRPC converted the translation differences between two point clouds in the spatial domain into the frequency domain and estimated the phase difference. The low-frequency components from the normalized cross-power spectrum were employed to obtain accurate 3D shifts by solving the parameters in the linear equation representing the phase difference angles using a robust estimator. The first step in the GRPC method is voxelization, which was achieved in this study by transferring the point clouds to 3D cubic signals that were subsequently transformed into the frequency domain using Fourier transformation. The correlated tensors of the two signals were obtained by using a 3D phase correlation. The correspondences between the phase-different angles and shift parameters were utilized to estimate the shift parameters by calculating the coefficient of the linear function between the phase-different angle and shift parameters. In the GRPC method, the noise in the signals and irrelevant parts were filtered out as high-frequency components, which made the method applicable to matching point clouds with low detailed correspondence.

For the correlation analysis, we compared the registration accuracy of the DAP and LiDAR point clouds of different matching scales, which were 2-$m$ fine registration achieved by using global matching and the iterative closest point (ICP) \citep{besl1992method} (denoted as h2f), 2-$m$, 3-$m$, 4-$m$, 5-$m$, and 10-$m$ coarse registration (denoted as h2c, h3c, h4c, h5c, and h10c, respectively), with reference to the measured tree heights. 
The tree height data of the field survey sample were extracted from the normalized DAP point cloud by combining the canopy boundary of single-tree segmentation \citep{xu2020tree}. 
Subsequently, the tree heights extracted from the point clouds were compared with the measured tree heights. A correlation analysis was performed. In the analysis, the linear model of the extracted tree height must fit a linear function whose slope was 1. The matching result that provided the highest correlation was used for further processing. Pearson's correlation coefficient was used as the evaluation metric and the calculation formula was as follows:
\begin{equation}
    r = \frac{N\sum x_iy_i-\sum x_iy_i}{\sqrt{n\sum x_i^2-(\sum x_i)^2}\sqrt{N\sum y_i^2-(\sum y_i)^2}},
\end{equation}
where $x$ denotes the measured tree height, and $y$ represents the height extracted from the DAP point cloud.
The optimal normalized DAP point cloud was obtained by comparing Pearson's correlation coefficients.

\subsection{Calculation of spectral and structural metrics}
We extracted the spectral information from multispectral imagery with five bands, namely blue, green, red, red, and near-infrared bands, with high spatial resolution. A set of vegetation indices was applied to the calculated spectral metrics according to the formula (see Table~\ref{tab:vegetation_indices}). The meanings and corresponding calculation formulas of the vegetation indices are listed in Table~\ref{tab:vegetation_indices}.

The structural information, which is also important in AGB estimation, was extracted from the normalized DAP point cloud using Fusion software and denoted as point cloud metrics. The point-cloud metrics used in this study are listed in Table~\ref{tab:structural_metrics}. 
\begin{table*}[ht!]
    \centering
    \caption{Summary of the spectral metrics with respective equations and references. It should be noted that spectral metrics were calculated based on raw imagery bands. \label{tab:vegetation_indices}}
    \resizebox{0.95\textwidth}{!}
    {\begin{tabular}{c c c}
    \toprule
       Spectral metric & Equation & Reference\\
       \midrule
       Atmospherically Resistant& $(p_{NIR}-p_{rb})/(p_{NIR}+p_{rb})$,&\multirow{2}{*}{\citep{kaufman1996strategy}}\\ 
       Vegetation Index (ARVI)&$p_{rb}=p_{red}-\gamma(p_{blue}-p_{red})$, $\gamma=0.5$\\
       Difference Vegetation Index (DVI) & $p_{NIR}-p_{red}$ & \citep{chang2010estimating}\\
       Green Normalized Difference & \multirow{2}{*}{$(p_{NIR}-p_{green})/(p_{NIR}+p_{green})$} & \multirow{2}{*}{\citep{verstraete1996designing}}\\
       Vegetation Index (GNDVI) & &\\
       Normalized Difference & \multirow{2}{*}{$(p_{NIR}-p_{red})/(p_{NIR}+p_{red})$} & \multirow{2}{*}{\citep{haboudane2004hyperspectral}}\\
       Vegetation Index (NDVI) & &\\
       Optimized Soil Adjusted & \multirow{2}{*}{$(p_{NIR}-p_{red})/(p_{NIR}+p_{red}+0.16)$} & \multirow{2}{*}{\citep{yuan2007model}}\\
       Vegetation Index (OSAVI)&&\\
       Red Green Ratio Index (RGRI) & $p_{red}-p_{green}$ & \citep{gamon1999assessing}\\
       Normalized Greenness (NormG) & $p_{green}/(p_{red}+p_{green}+p_{green})$ & \citep{fraser2017calibrating}\\
        \bottomrule
    \end{tabular}}
\end{table*}

\begin{table*}[ht!]
    \centering
    \caption{Summary of the structural metrics extracted from the DAP point cloud.\label{tab:structural_metrics}}
    \resizebox{0.9\textwidth}{!}
    {\begin{tabular}{c c}
    \toprule
       Metrics & Description\\
       \midrule
        \multirow{2}{*}{Percentile heights ($h_{25},h_{50},h_{75},h_{95}$)} & Percentiles of the canopy \\
        & height distributions (25th, 50th, 75th, 95th)\\
        Mean height ($h_{mean}$) & Average height of non-ground\\
        \multirow{2}{*}{Coefficient of variations in height cover ($h_{cv}$)} & Coefficient of variation in heights of \\
        & Non-ground coverage over two meters\\
        \bottomrule
    \end{tabular}}
\end{table*}

\subsection{Calculation of AGB-tree estimates of sample plots} \label{sec:individual_estimation}
To develop a model for plot-level AGB estimation, we first calculated the AGB for each sample plot using an individual-tree-based method as our observed AGBs.
Here, the biomass (ton, $t$) was estimated using the tree-specie-based function of the DBH (m, $m$) and tree height (m, $m$), which was the maximum height of the canopy.
The DBH was calculated using the height-DBH model based on the tree height and estimated tree species. The height-DBH models were obtained by fitting the tree height to the DBH curve based on the diameters and heights of 502 measured trees in the field.
The calculation formulas for the height-DBH conversion and AGB models are provided in Table~\ref{tab:individualtree_model}.

To accurately estimate each sample plot, individual tree crowns were first delineated by applying an advanced multiresolution segmentation algorithm to the composited CHM and RGB imagery. 
Subsequently, the individual tree species could be classified using spectral and point-cloud metrics extracted from the multispectral imagery and CHM \citep{xu2020tree}.
Once the tree crowns were segmented, we extracted the tree height for each individual tree crown from the CHM.
Finally, based on the tree species and estimated tree heights, the AGB-tree estimates were calculated by summarizing the AGB of each individual tree.

\begin{table*}[ht!]
    \centering
    \caption{Tree height-DBH models and AGB models for different tree species \citep{tree2020guideline}. $W_S$, $W_B$, $W_L$, $W_P$, and $W_T$ denote stem biomass, leaf biomass, branch biomass, pod biomass, and tree biomass, respectively. \label{tab:individualtree_model}}
    \resizebox{\textwidth}{!}
    {\begin{tabular}{c c c}
    \toprule
    Tree species & Height-DBH models & AGB calculation formula\\
    \midrule
    \multirow{3}{*}{A.N.} & & $W_S = 0.02739(D^2H)^{0.898869}$; $W_B=0.01497(D^2H)^{0.875639}$;\\
    & $D=0.0091H^2+$ & $W_L = 0.01059(D^2H)^{0.66681}$; $W_P=0.0121(D^2H)^{0.854295}$;\\
    & $1.1147H+7.0082$ & $W_T=W_S+W_B+W_L+W_P$\\
    & & \\
    
    \multirow{3}{*}{Q.A.} & \multirow{3}{*}{$D=0.5289H^{0.9858}$} & $W_S=0.02739(D^2H)^{0.898869}$; $W_B=0.01497(D^2H)^{0.875639}$;\\
    & & $W_L = 0.01059(D^2H)^{0.66681}$; $W_P=0.0121(D^2H)^{0.854295}$;\\
    & & $W_T=W_S+W_B+W_L+W_P$\\
    & & \\
    
    \multirow{3}{*}{C.O.} & \multirow{3}{*}{$D=0.8169H^{1.2791}$} & $W_S=0.3507(D-1.1948)^2$; $W_B=0.03017D^{2.3643}+0.051$;\\
    & & $W_L = 0.01813D^2-0.2477$;\\
    & & $W_T=W_S+W_B+W_L$\\
    & & \\
    
    \multirow{3}{*}{A.F.} & \multirow{3}{*}{$D=0.3399H^{1.6976}$} & $W_S=0.02739(D^2H)^{0.898869}$; $W_B=2.6259+0.0633D^2$;\\
    & & $W_L = 3.5207\times10^{-4}(15.9739+D)^3$; $W_P=0.054124(D-3.502)^2$;\\
    & & $W_T=W_S+W_B+W_L+W_P$\\
    & & \\
    
    \multirow{3}{*}{T.C.} & \multirow{3}{*}{$D=1.8482H^{1.067}$} &
    $W_S = 0.3274(D-3.6998)^2$; $W_B=0.01497(D^2H)^{0.875639}$;\\
    & & $W_L=0.01059(D^2H)^{0.66681}$; $W_P=0.0121(D^2H)^{0.854295}$;\\
    & & $W_T=W_S+W_B+W_L+W_P$\\
    & & \\
    
    P.D. & $D=1.5542H^{1.0014}$ & $W_T=0.07052(D^2H)^{0.9381716}$\\
    & & \\
    
    \multirow{3}{*}{T.C.} & \multirow{3}{*}{$D=11.666e^{0.0544H}$} & $W_S=0.0105(D^2H)^{1.0652}$; $W_B=0.8775(D^2H)^{0.9894}$;\\
    & & $W_L=0.033(D^2H)^{0.9352}$; $W_P=0.043(D^2H)^{0.6628}$;\\
    & & $W_T=W_S+W_B+W_L+W_P$\\
    & & \\
    
    \multirow{3}{*}{P.A.} & \multirow{3}{*}{$D=3.5113H^{0.8022}$} & $W_S=0.02091(D^2H)^{0.9285}$; $W_B=0.1336(D^2H)^{0.8870}$;\\
    & & $W_L=0.007974(D^2H)^{0.8998}$; $W_P=0.011332(D^2H)^{0.9285}$;\\
    & & $W_T=W_S+W_B+W_L+W_P$\\
        \bottomrule
    \end{tabular}}
\end{table*}

\subsection{Predictive model for AGB-plot estimates} \label{sec:development_model}
Using the AGB-tree estimates of the sample plots as our observed AGB, we aimed to develop a predictive AGB model based on the spectral and point cloud metrics. 
The best metrics and the subset of metrics for modeling the AGB-plot estimates were determined through an all-subset regression and correlation analysis, respectively. 
Here, all the metrics were grouped into four subsets: (i) point cloud metrics derived from the DAP point clouds of coarse normalization (seven metrics); (ii) point cloud metrics derived from DAP point clouds of fine normalization (seven metrics); (iii) multispectral metrics (nine metrics); and (iv) multispectral and point cloud metrics derived from the DAP point cloud of fine normalization (16 metrics). 
Correlation analysis was conducted by comparing three key evaluation metrics: coefficient of determination ($R^2$), root mean square error (RMSE), and relative root mean square error ($rRMSE$). 
The equations are as follows:
\begin{equation}
    R^2 = 1-\frac{\sum^n_{i=1}(x_i-\hat{x}_i)^2}{\sum^n_{i=1}(x_i-\Bar{x}_i)^2},
\end{equation}
\begin{equation}
    RMSE = \sqrt{\frac{1}{n}\sum^n_{i=1}(x_i-\hat{x}_i)^2},
\end{equation}
\begin{equation}
    rRMSE = \frac{RMSE}{\Bar{X}}\times 100{\%},
\end{equation}
where $x_i$ is the actual value of the forest index, $\Bar{x}_i$ denotes the mean value of the forest index, $\hat{x}_i$ denotes the predicted value of the forest index, $n$ is the number of sample plots, and $i$ is the number of variables.

Finally, the optimal predictive model of the AGB-plot estimates was applied to the study area to achieve the wall-to-wall biomass mapping of large-scale forests. The resolution of the final mapping of the AGB in the study area was 15 m.

\section{Results}
\subsection{Selection of the optimal registration result between DAP and LiDAR point clouds}\label{sec:results_registration}
\begin{figure}[htb]
    \centering
    \includegraphics[width=\textwidth]{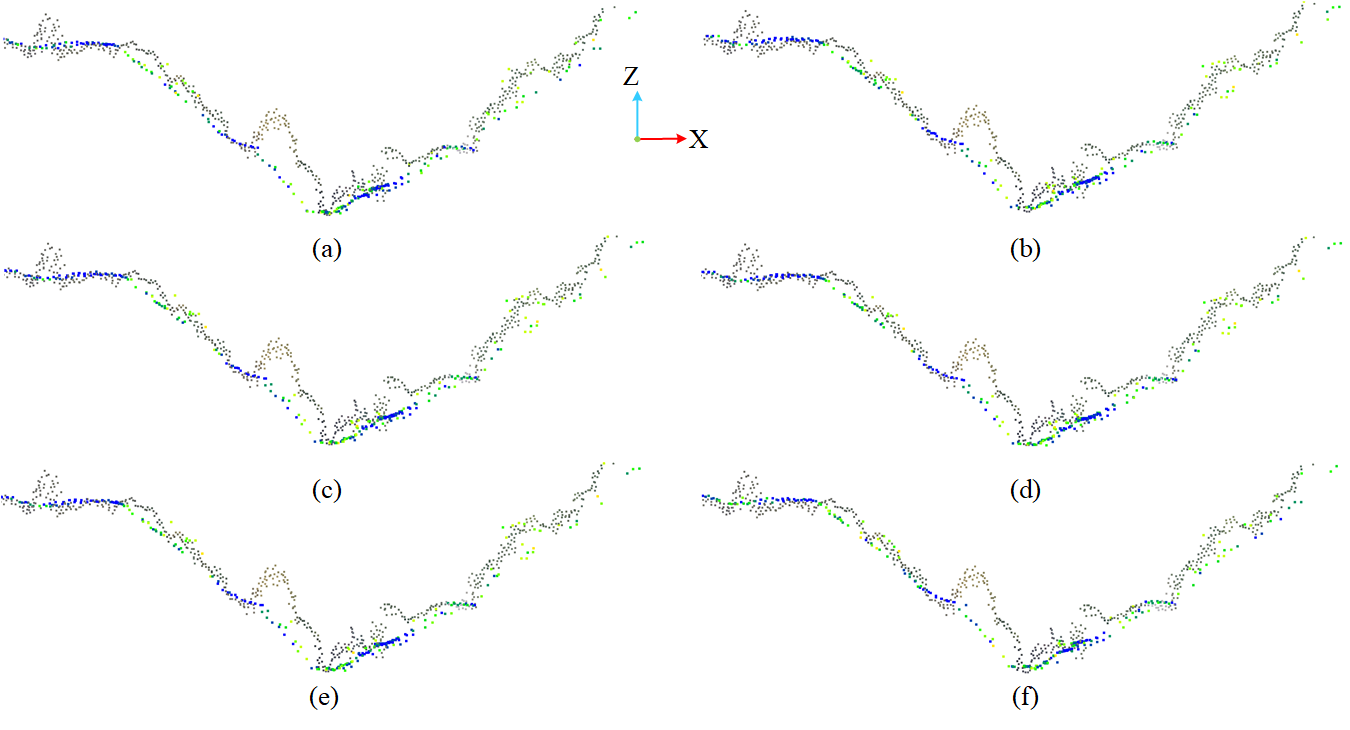}
    \caption{Cross-sections of registration results. a) 2-$m$ fine registration using global registration followed by ICP; b) 2$m$ coarse registration; c) 3-$m$ coarse registration; d) 4-$m$ coarse registration; e) 5-$m$ coarse registration; f) 10-$m$ coarse registration. The DAP point cloud is colored in RGB and the LiDAR point clouds are colored in intensity.}
    \label{fig:registration_results}
\end{figure}

\begin{figure}[ht!]
    \centering
    \includegraphics[width=\textwidth]{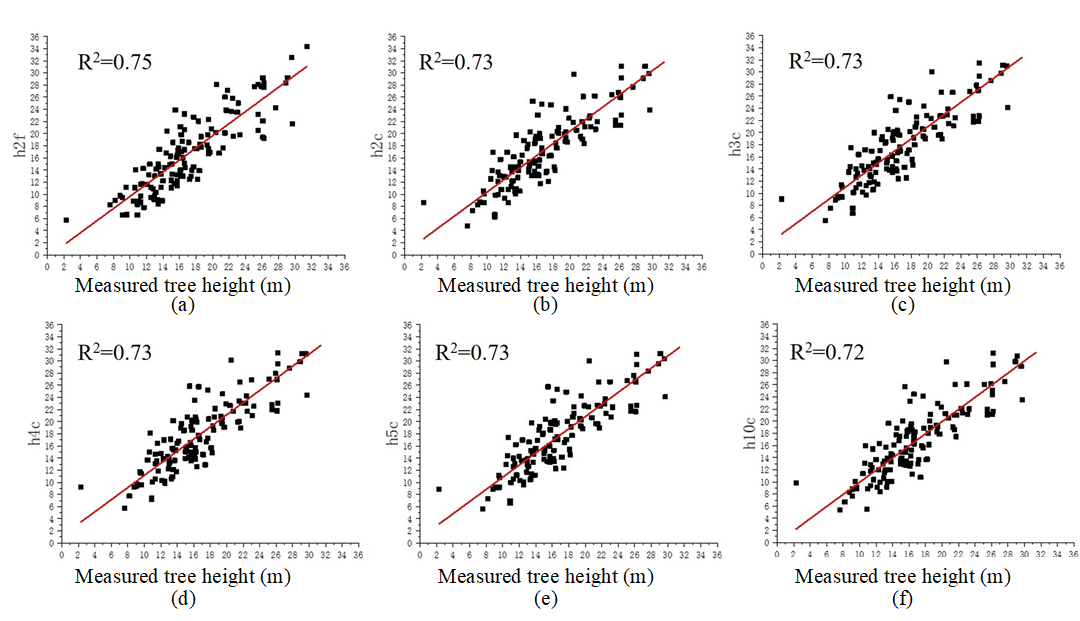}
    \caption{Scatterplot of the measured tree heights and tree heights extracted based on different registration results, namely a) 2-$m$ fine registration achieved by global registration followed by ICP, b) 2-$m$ coarse registration, c) 3-$m$ coarse registration, d) 4-$m$ coarse registration, e) 5-$m$ coarse registration, and f) 10-$m$ coarse registration.}
    \label{fig:tree_height}
\end{figure}

The key to accurately estimating tree heights lies in achieving an optimal registration between the DAP and LiDAR point clouds. The registration results are presented in Fig.~\ref{fig:registration_results}. As shown in the figure, the point clouds are accurately matched after registration using different parameterizations from a general perspective. A correlation analysis was conducted for further evaluations of the registration results. The results of the correlation analysis between the extracted tree heights based on registration results under different parameterizations and the measured tree heights are provided in Table~\ref{tab:matching_accuracy} and Fig.~\ref{fig:tree_height}. As shown in the table, for the 502 field-measured trees, the extracted tree heights and measured tree heights are strongly correlated, with Pearson's correlation coefficient varying from 0.85 to 0.86 for all the registration results. No significant bias is observed when using the normalized DAP point clouds to measure tree heights. However, the correlation coefficient is inversely proportional to the matching scale. The measurement of the tree heights based on the point clouds is the most accurate when coarse registration is performed using the smallest matching scale (2 m), followed by fine registration using the ICP. Moreover, as depicted in the scatterplots in Fig~\ref{fig:tree_height}, the tree measurements based on the combination of the DAP and LiDAR point clouds are more accurate when the trees are lower with a smaller standard deviation. However, when measuring the height of taller trees using a combination of point clouds, we had fewer measured samples and stronger bias.
\begin{table*}[ht!]
    \centering
    \caption{Comparison of registration accuracy under different parameterization with respect to measured tree heights.\label{tab:matching_accuracy}}
    \resizebox{0.8\textwidth}{!}
    {\begin{tabular}{c c c c c c c}
    \toprule
       Equation & \multicolumn{6}{c}{$y=a+b*x$}\\
       \midrule
        model & h2f & h2c & h3c & h4c & h5c & h10c\\
        RMSE & 1273.153 & 1268.057 & 1243.952 & 1239.448 & 1251.618 & 1348.068\\
        Pearson's r & 0.864 & 0.857 & 0.858 & 0.858 & 0.858 & 0.849\\
        $R^2$ & 0.746 & 0.733 & 0.734 & 0.733 & 0.735 & 0.720\\
        \bottomrule
    \end{tabular}}
\end{table*}

\subsection{Estimation of AGB from individual trees}
\begin{figure}[ht!]
    \centering
    \includegraphics[width=\textwidth]{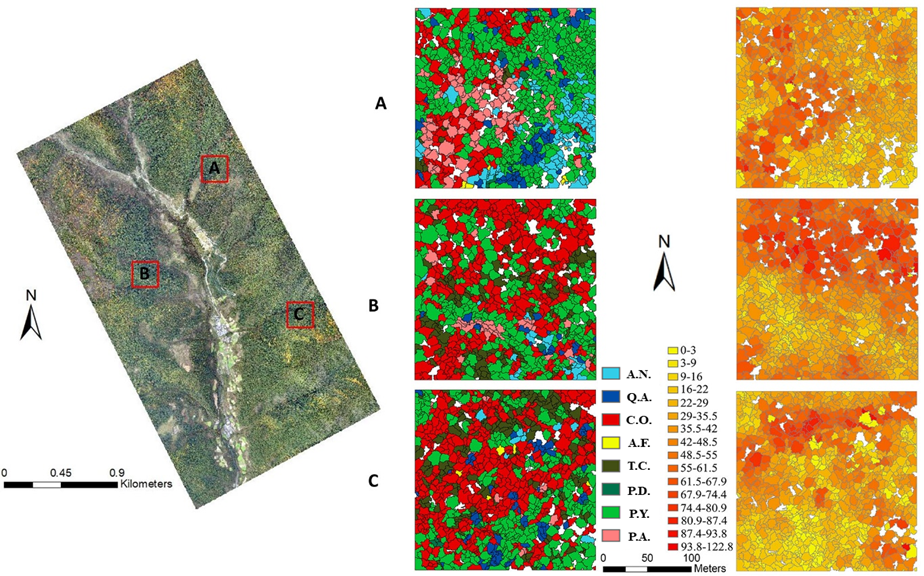}
    \caption{Results of the individual-tree-based AGB estimation of three sample plots for illustration. The area of each sample plot is 200 $m$ $\times$ 200 $m$.}
    \label{fig:results_individual}
\end{figure}

The results of the species classification and biomass estimation at the individual tree level are shown in Fig.~\ref{fig:results_individual}. Three sample plots were selected to illustrate the results of the individual tree crown delineation and AGB estimation. As shown in the figure, individual trees with clear crown boundaries were detected. These results revealed that P.Y. was the most widely distributed tree species in the selected samples of the study area. However, other typical broad-leaf species (A.F.) had few samples in the selected areas.   Fig.~\ref{fig:results_individual} also shows the results of biomass mapping based on individual trees. Overall, the AGB of the tree-based estimates varies from 0 to 122 $t/h$. The distribution of the biomass does not exhibit a positive relationship with the distribution of tree species; rather, it is concentrated in some spatial areas with dense trees and large crown sizes. For the dominant tree species, both the tree crowns and the estimated AGB vary significantly, indicating that the PYs have rich age levels and a wide range of tree heights in the study area.

\subsection{Inversion of plot-based AGB estimation model} \label{sec:results_development}
\begin{figure}[ht!]
    \centering
    \includegraphics[width=0.9\textwidth]{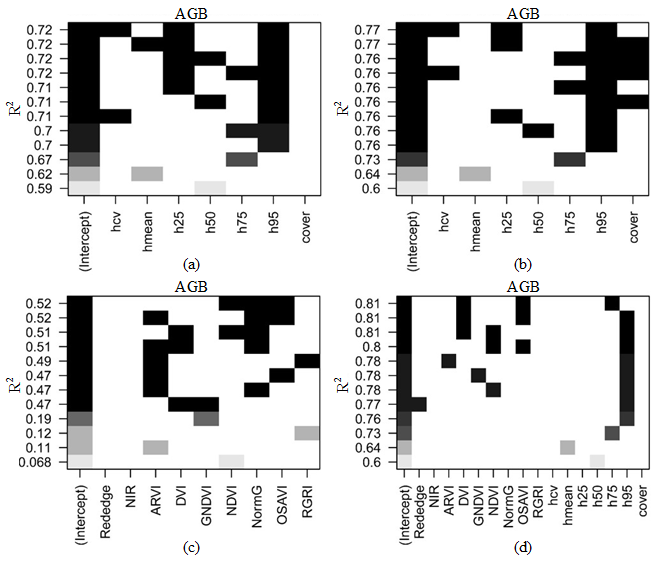}
    \caption{Select the optimal metrics as the inputs for AGB modeling using all subsets regression. a) Point cloud metrics extracted from DAP point cloud of coarse normalization, b) point cloud metrics extracted from DAP point cloud of fine normalization, c) multispectral metrics, and d) multispectral and point cloud metrics extracted from DAP point cloud of fine normalization.}
    \label{fig:results_subset}
\end{figure}

The linear model was utilized to invert the plot-based AGB estimation model. We selected the best independent variables using all subsets regression. Fig.~\ref{fig:results_subset} displays a regression that considers both spectral and structural metrics. The accuracy was assessed based on 34 sample plots distributed across the entire study site. The optimal subsets of metrics were selected for the four different data inputs. Although almost all the metrics contributed to the modeling of the AGB, the significance of the different metrics differed. For the subset of the point cloud metrics extracted from the DAP point cloud of coarse normalization, the model fitness $R^2$ metrics varied from 0.59 to 0.72, and the optimal metrics were obtained as $h_{cv}$, $h_{25}$, and $h_{95}$. For the DAP metrics of fine normalization, $h_{cv}$, $h_{25}$, and $h_{95}$ produced the highest $R^2$ and were thus selected as the inputs for the linear regression of the AGB. The $R^2$ metric was in the 0.068-0.52 range for all response variables of the multispectral metrics and NDVI, NormG, and OSAVI exhibited the highest accuracies. For the combination of multispectral and DAP point clouds of fine normalization, $R^2$ varied from 0.6 to 0.81 and OSAVI, DVI, and $h_{75}$ were selected as the optimal subset.
The optimal subsets for each type of input data were then utilized to develop a linear regression model.

\begin{table*}[ht!]
    \centering
    \caption{Description of AGB predictive models and accuracy assessment of each models.\label{tab:agb_models}}
    \resizebox{\textwidth}{!}
    {\begin{tabular}{c c c c c}
    \toprule
       Data & AGB Predictive Models & $R^2$ & $RMSE$ & $rRMSE$\\
       \midrule
        \multirow{2}{*}{DAP{\_}coarse} & $W_{plot}=3.851\times h_{25}+1.945\times h_{95}+$ & \multirow{2}{*}{0.72} & \multirow{2}{*}{14.57} & \multirow{2}{*}{11.81}\\
        & $53.484\times h_{cv}+44.215$ & \\
        \multirow{2}{*}{DAP{\_}fine} & $W_{plot}=3.347\times h_{25}+2.154\times h_{95}+$ & \multirow{2}{*}{0.77} & \multirow{2}{*}{13.30} & \multirow{2}{*}{10.78}\\
        & $66.913\times hcv+33.315$ & \\
        \multirow{2}{*}{Multi-spectral} & $W_{plot}=660.22\times NDVI-656.06\times NormG-$ & \multirow{2}{*}{0.52} & \multirow{2}{*}{19.08} & \multirow{2}{*}{15.46}\\
        & $836.03\times OSAVI+85.11$ &\\
        \multirow{2}{*}{Multi-spectral and DAP{\_}fine} & $W_{plot}=4398\times OSAVI-17410\times DVI+$ & \multirow{2}{*}{0.81} & \multirow{2}{*}{11.99} & \multirow{2}{*}{9.71}\\
        & $4.018\times h_{75}-57.78$ &\\
        
        \bottomrule
    \end{tabular}}
\end{table*}
Table~\ref{tab:agb_models} presents the results of the developed AGB predictive models and their corresponding accuracy assessment. As shown in the table, the $R^2$ metrics varies from 0.72 to 0.81 with different types of data inputs. Compared with the model utilizing coarsely normalized DAP point clouds, the $R^2$ and $RMSE$ are improved by 7{\%} and 9{\%} when finely registered DAP point clouds are used for estimating the AGB. The accuracy of the AGB estimates is improved by 9{\%} and 5{\%} in terms of the $RMSE$ and $R^2$, respectively, when spectral metrics are added to only point cloud structural metrics. In addition, compared with results obtained by using only spectral metrics, the results of the combination of spectral and point cloud structural metrics are enhanced by 37{\%} and 55{\%} in terms of the $RMSE$ and $R^2$, respectively.

\begin{figure}[ht!]
    \centering
    \includegraphics[width=0.8\textwidth]{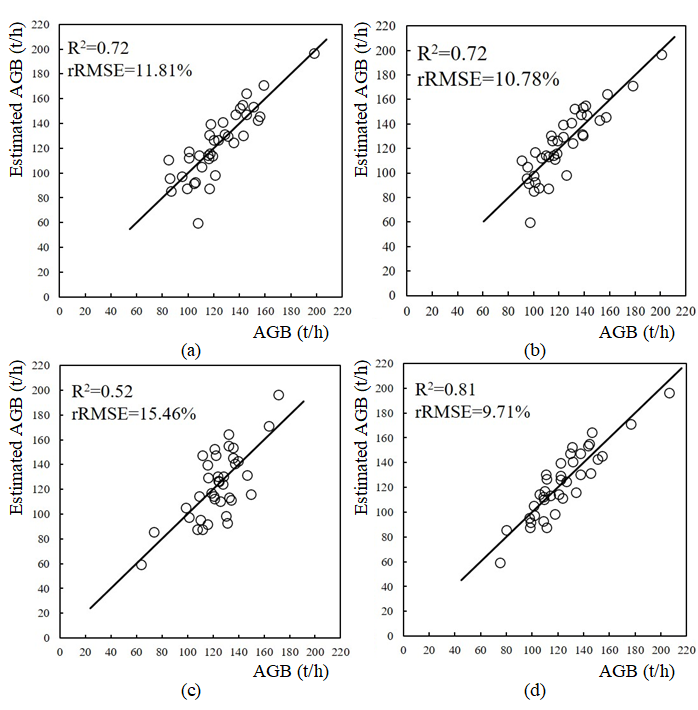}
    \caption{Scatterplot of the calculated results of biomass of sample plots in each model. a) Point cloud metrics extracted from the DAP point cloud of coarse normalization, b) point cloud metrics extracted from DAP point cloud of fine normalization, c) multispectral metrics, and d) multispectral and point cloud metrics extracted from DAP point cloud of fine normalization.}
    \label{fig:results_scatterplot}
\end{figure}

In Fig.~\ref{fig:results_scatterplot}, we present the corresponding scatterplots of the calculated results with the predicted values as the response variables.
A correlation analysis of the models demonstrated that the relationship between the DAP measures and predicted AGB was close to 1:1. The multispectral and DAP models had the highest accuracy, indicating that the point cloud and multispectral metrics had desirable synergistic effects in the prediction of AGB. The scatterplot results indicated that the data source had a significant impact on the model's performance, with the $R^2$ varying from 0.52 to 0.61. As shown by the results, the linear model was more effective for the DAP point clouds than for multispectral data. The combination of the DAP point cloud and multispectral data improved the performance of the models across all the regression methods by reducing the $RMSE$ and increasing the $R^2$.
Finally, the optimal linear model used for predicting the plot-level AGB was obtained as follows.
\begin{equation}
    AGB = 4398\times OSAVI-17410\times DVI+4.018\times h_{75}-57.78
\end{equation}
All the parameters played an important role in the AGB modeling.

\subsection{Landscape plot-based AGB estimation}
\begin{figure}[ht!]
    \centering
    \includegraphics[width=0.95\textwidth]{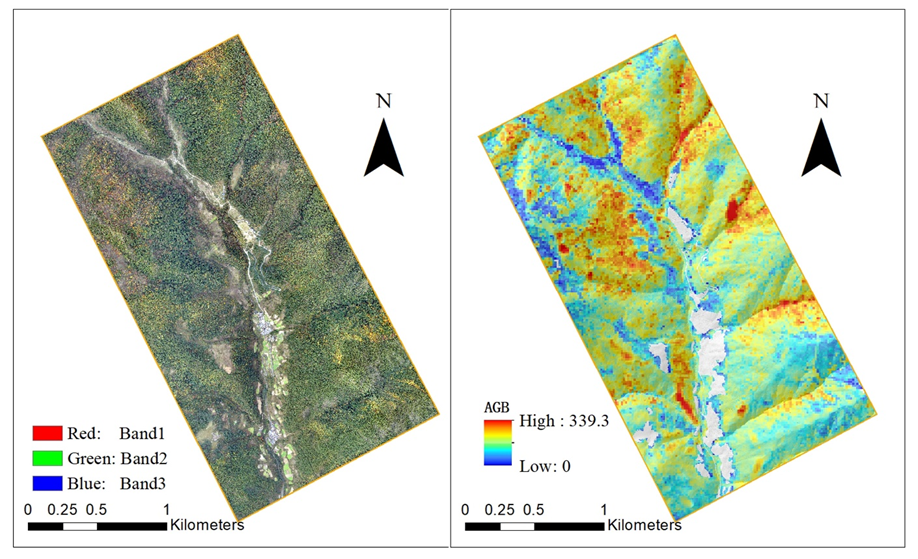}
    \caption{Landscape biomass mapping in the study area. The optimal model is substituted for biomass mapping.}
    \label{fig:biomass_map}
\end{figure}

The landscape mapping of the AGB of the natural reserves is shown in Fig.~\ref{fig:biomass_map}. As shown in the figure, most of the study area was covered by dense trees with high AGBs. In general, the AGBs of the study area varied from 0 to 350 $t/hm$. The spatial pattern of the AGBs followed broad patterns of forest cover, which also showed an obvious response to the distribution of human living areas. The topography of the study area also showed a relationship with the distribution of the landscape AGBs. The AGBs of valleys was different from those of ridges. In valleys, tree coverage was less, whereas, in ridges, denser canopy coverage with higher AGBs existed. This may be attributable to the difference in the distributions of tree species owing to variations in illumination and humidity. Moreover, there are wide height variations in the natural reserve, which could also be a major cause of the uneven distribution of vegetation.

\section{Discussion}
\subsection{Influence of registration between DAP and LiDAR point clouds on measuring tree heights}
As tree height is an important indicator in AGB estimation, precisely extracting canopy heights from the acquired data is vital. In the synergistic system, two data sources were utilized to extract canopy heights; thus, the registration became the key processing step in the whole workflow for biomass estimation. In our study, we evaluated the registration between the DAP and LiDAR point clouds with varying parameterizations and successfully confirmed that the accuracy of the registration results and, by extension, the height extraction precision, varied according to the parameterizations of the applied registration method. In general, fine registration yielded higher registration accuracy and improved the precision of the estimated heights. As stated in Section~\ref{sec:results_registration}, the registration results had less impact on the measurement of the heights of middle-sized trees. However, for tall trees, the extraction of canopy height was more influential. Thus, when a large area with strong variations in tree heights was considered, precise registration between the DAP and LiDAR point clouds ensured precise extraction of heights of trees, as distinct from tree species, ages, and growth states, and was key for further extraction of other basic forest parameters such as DBHs and AGBs.
\subsection{Improvement of predictive model through integration of DAP and LiDAR data}
Ecological questions in the study area can be addressed and analyzed based on the estimated biomass of the landscape.  
A key novelty of this study lies is the optimization of cost efficiency through the complementary fusion of multispectral images and sparse LiDAR point clouds obtained using a synergistic system. The proposed framework is an efficient approach for obtaining wall-to-wall estimates of mountainous forest structures. In Section~\ref{sec:results_development}, we developed predictive models using only DAP metrics, spectral metrics, and a combination of both spectral and point-cloud metrics. 

First, our study confirmed the reliability of point cloud structural-metrics-based prediction in forested areas, consistent with the findings of studies using individual tree-based methods.
Airborne LiDAR data can provide measurements of the terrain, which serve as an important input for obtaining accurate tree heights from the normalized DAP point cloud. In our study, we relied on a registration method that matches both point clouds and normalized DAP point clouds using terrain information provided by LiDAR point clouds. The registration accuracy affected the estimation of normalized heights, thus influencing the extraction of reliable structural metrics. Accurate registration of DAP and LiDAR point clouds was essential for further model prediction and biomass mapping.

Second, the normalized DAP point clouds share similar characteristics when producing reliable CHM, which is one of the most important inputs for calculating the structural metrics for predicting AGB. As mentioned in Section~\ref{sec:results_development}, the structural metrics provided by the normalized DAP point cloud significantly influenced the predictive model because it provided a valuable variable that helped in predicting biomass under tree crowns, especially in densely distributed forest areas. Although the spectral attributes indicated the distribution of different tree species, without point cloud structural metrics, forest biomass is still underestimated.  

\subsection{Comparison of individual-tree based and plot-based estimates of AGB}
Our study supports the development of plot-based approaches for estimating landscape biomass in forested areas. Compared to the individual tree-based method, the plot-based methods have distinct advantages owing to their simplicity, scalability to large landscapes, and ability to adjust the predictive model to local conditions based on tree and carbon density. The plot-based method can provide continuous coverage of AGB estimates in forested areas. In addition, the spatial resolution can be changed to estimate the AGB for different practical purposes. However, the plot-based method has certain limitations. In areas with strongly heterogeneous canopies, the use of plot-based methods for estimating the AGB is not suitable. For instance, in areas near a village in a natural reserve, there are only sparsely distributed trees with low canopy cover. Spectral and structural metrics are not suitable for estimating biomass in this case. 

In the individual-tree-based method, individual trees can be detected using tree canopy segmentation algorithms. This can provide a highly accurate estimation of biomass based on single-tree detection results. This is also the reason we utilized the individual-tree-based estimates of the AGB as support data for developing an aggregated plot-based model for difficult areas when there is a significant amount of fieldwork. However, individual tree crown delineation and classification of tree species will significantly increase the workload when estimating the biomass of large-scale forest areas. The necessity of this for individual tree detection should be determined. Thus, when mapping a large-scale forest area, the plot-based method can accurately measure the biomass in closed canopy-covered areas. However, at the border of the forest or buffer areas with human-living areas, the characteristics of tree distributions should be further considered.

\subsection{Importance of selecting optimal spectral and point cloud structural metrics}
Determining the best metrics is essential for estimating biomass. As discussed in Section~\ref{sec:results_development}, the results for different subsets of predictor variables indicate that the data inputs have a significant impact on the performance of the predictive model. First, for the point cloud metrics, the predictor variables indicating percentile heights and variation in heights play an important role in estimating the AGBs. We can define the growth states of trees using these tree height-related metrics. In addition, as explained in Section~\ref{sec:individual_estimation}, tree heights are directly related to DBH and, thus, have a significant influence on the AGB models of individual trees. This characteristic can also be observed when using the plot-based method. This indicates the importance of extracting the precise tree height during the biomass estimation process.

In addition to the height distribution patterns indicated by the structural information, spectral metrics are crucial for mapping AGBs with high accuracy. The results revealed that the spectral indices indicating the soil and vegetation attributes contributed more to the modeling of the AGB based on spectral metrics.

Combining selected spectral and point cloud structural metrics improved the accuracy of the estimation of the AGB. Attributes from different data sources were aggregated to generate a plot-based predictive model with less $RMSE$, compared with the reference data provided by the individual tree estimation results. 
\subsection{Challenges of large-scale forest biomass mapping}
Based on the former results and analysis, the main considerations and possible insights for mapping large-scale forest biomass, especially in mountainous plateau areas, are as follows:
\begin{itemize}
    \item Availability of multispectral and LiDAR data
    \item Characteristics of measured areas
    \item Selection of appropriate variables
    \item Effect of LiDAR strip overlaps and adjustment
    \item Framework for data processing
\end{itemize}

The availability of spectrally and spatially comprehensive data is essential for large-scale mapping tasks. Robust and accurate measurement of forest areas is crucial for ensuring an accurate estimation of the AGB in the region of interest. However, because of the limitations of observation conditions, how to acquire data under specific conditions is a crucial research topic. In this study, as the region of interest was located in a high-altitude and mountainous area, utilizing common data acquisition methods, such as nap-of-the-earth LiDAR techniques, was difficult. This is also the reason why we combined a UAS-based LiDAR and a camera to obtain dense point clouds.  

The characteristics of the measured areas not only limited the choice of field work methods but also determined the method applied for biomass mapping. As mentioned earlier, the altitude of the forest and topography of the corresponding region limited data acquisition. Furthermore, the characteristics of the measured areas, such as the conditions of canopy cover and distributions of residential areas, influenced the development of the AGB estimation method. For instance, the plot-based method would be effective for estimating the AGB in homogeneous areas such as dense canopy-covered areas. However, for forest borders or buffer areas connected to residential areas or farms, it would be better to estimate AGB using single trees because the measured area is comparatively heterogeneous.

In addition to the selection of the strategy utilized for estimating AGB based on the characteristics of the measured area, the selection of the attributes to be included in the mapping process is important. The subset of the input variables of the model must capture the major characteristics of the forest. They should also be variables that are harmonious across regions of interest in both model generation and mapping tasks. However, these variables should be robust to noise and variations. In our study, we tested the spectral and point cloud metrics derived from the multisource data to ensure that we selected the optimal attributes among them.

Although the registration method applied in the framework of data processing can handle matching point clouds with low overlapping ratios, sufficient overlaps are still required, particularly for the overlaps between LiDAR strips. Small overlaps can decrease the number of flights and, thus, reduce the flight cost; however, they decrease the quality of the integrated LiDAR point clouds based on the limitations of the matching algorithms. The poor quality of the integrated LiDAR point clouds leads to further registration errors in the matching of different data sources. Overall, the overlapping ratios between LiDAR strips should be carefully selected in real applications of the framework of the information fusion strategy for biomass mapping.

Mapping large-scale forest areas requires large amounts of multisource geospatial data. Therefore, in this study, developing an optimal framework for processing these data was a key preoccupation toward verifying the success and efficiency of accomplishing the mapping task. As previously mentioned, the tree-based method may provide comparatively precise AGB estimates. However, for large-scale mapping, computational limitations and memory issues must also be considered.

\section{Conclusions}
In this study, we explored the potential of integrating UAS-based multispectral images and LiDAR point clouds to estimate forest biomass in a plateau mountainous reserve with high relief and undulated terrain. We deployed a hierarchical strategy combining an information fusion approach and a plot-based regression method based on selected spectral and point cloud metrics. The findings of the study may be summed up as follows:
\begin{itemize}
    \item Although sparse LiDAR point cloud acquired at a high flight speed could not be used alone to estimate the AGB owing to its low point density, it could serve as an important data source for terrain estimation and as support data for accurate estimation of tree heights, thus, significantly contributing to the estimation of AGB.
    \item In the entire workflow of utilizing DAP and LiDAR point clouds for AGB estimation, registration of these two data sources played an important role, as the registration results determined the extraction of CHM. Fine registration produced a higher registration accuracy and improved the precision of the estimated tree height.
    \item The accuracy of the AGB estimates could be improved by 9{\%} and 5{\%} in terms of the $RMSE$ and $R^2$, respectively, by adding spectral metrics to point cloud structural metrics. Compared with results obtained using only spectral metrics, the results obtained using the combination of spectral and point cloud metrics were better by 37{\%} in terms of the $RMSE$ and 55{\%} in $R^2$. The most important metrics when combining both data sources were OSAVI, DVI, and $h_{75}$, an aggregation of structural attributes such as tree height, height percentile, and spectral attributes of vegetation and soil.
    \item The synergistic UAS system that rapidly acquired both optical imagery and LiDAR point clouds with low cost and risk can be utilized to accurately estimate the AGB in challenging forest areas of undulated terrain. Both individual- and plot-based AGB estimation methods can be applied when combining two data sources.
\end{itemize}
This study contributes to the investigation of the potential of utilizing a synergistic system that combines optical and LiDAR sensors and the proposed workflow to estimate the AGB of a subtropical forest in a plateau mountainous reserve. The forest biomass can be estimated more accurately by developing an AGB prediction strategy based on the characteristics of the measured areas. 
\section{Acknowledgement}

This work was supported by the National Natural Science Foundation of China (Project No. 42171361), by the Research Grants Council of the Hong Kong Special Administrative Region, China, under Project PolyU 25211819, and partially by the Hong Kong Polytechnical University under Projects 1-ZE8E and 1-ZVN6.





\bibliographystyle{elsarticle-harv}
\bibliography{elsarticle}

\end{document}